\documentclass{article}
    \PassOptionsToPackage{numbers, compress}{natbib}
    \usepackage[final]{neurips_2023}

\usepackage[utf8]{inputenc} % allow utf-8 input
\usepackage[T1]{fontenc}    % use 8-bit T1 fonts
\usepackage{hyperref}       % hyperlinks
\usepackage{url}            % simple URL typesetting
\usepackage{booktabs}       % professional-quality tables
\usepackage{amsfonts}       % blackboard math symbols
\usepackage{nicefrac}       % compact symbols for 1/2, etc.
\usepackage{microtype}      % microtypography
\usepackage{xcolor}         % colors
\usepackage{amsmath}
\usepackage{amsthm}
\usepackage{enumitem}
\usepackage{mdframed}
\usepackage{graphicx}
\usepackage{adjustbox}
\usepackage{cleveref}
\usepackage{setspace}
\usepackage{subcaption} % Required for subfigures

\theoremstyle{definition}
\newtheorem{example}{Example}
\title{
Data Ambiguity Strikes Back: How Documentation Improves GPT's Text-to-SQL
}

\author{
  Zezhou Huang \\
  Columbia University\\
  \texttt{zh2408@columbia.edu} \\
  \And
  Pavan Kalyan Damalapati \\
  Columbia University\\
  \texttt{pd2720@columbia.edu} \\
  \AND
  Eugene Wu \\
  DSI, Columbia University\\
  \texttt{ewu@cs.columbia.edu} \\
}

\begin{document}

\maketitle
% \vspace{-18mm}
\begin{abstract}
Text-to-SQL allows experts to use databases without in-depth knowledge of them. However, real-world tasks have both query and data ambiguities.
Most works on Text-to-SQL focused on query ambiguities and designed chat interfaces for experts to provide clarifications.
In contrast, the data management community
has long studied data ambiguities, but mainly addresses error detection and correction, 
rather than documenting them for disambiguation in data tasks. 
This work 
delves into these data ambiguities in real-world datasets. 
We have identified prevalent data ambiguities of value consistency, data coverage, and data granularity that affect tasks. We examine how documentation, originally made to help humans to disambiguate data, can help GPT-4 with Text-to-SQL tasks. By offering documentation on these, we found GPT-4's performance improved by  $28.9\%$.

\end{abstract}

\section{Introduction}

Text-to-SQL is widely used as it allows domain experts who aren't familiar with database structures or SQL to access data. 
Although specialized models have been developed and show promising results~\cite{yu2019cosql,veltri2023data},
recent studies have found that, by increasing the size of both the model and training data, general-purpose Large Language Models (LLMs) like GPT-4 with around 1.7T parameters can achieve state-of-the-art performance~\cite{hagleitner2023gpt4sql,liu2023comprehensive,dong2023c3} in Text-to-SQL tasks using  the Spider benchmarks~\cite{yu2018spider}.

Unlike the Spider dataset, which is characterized by its well-structured schema and clean data, real-world Text-to-SQL tasks  often present challenges due to ambiguities both from query and data:

\begin{itemize}[leftmargin=0.5cm]
\item  {\bf Query Ambiguity}: The queries provided by domain experts can be interpreted in multiple ways with respect to the data. Common query ambiguities include query term  holding multiple meanings~\cite{wang2022know,yu2019cosql,veltri2023data,yu2019cosql,zhang2020did}, or the output schema being under-specified~\cite{dong2023c3, zhang2023towards}.
\item {\bf Data Ambiguity}: The real-world concepts encapsulated within the data can be interpreted differently. 
Data ambiguity is a fundamental aspect of data, independent of  the queries or the Text-to-SQL tasks at hand, and has been studied  by the data management 
community for decades.
This paper follows the scope of data ambiguities established by  previous works ~\cite{batini2009methodologies,geareview}, which includes value consistency~\cite{madnick2009overview,raman2001potter} (e.g., do the values follow consistent formats?), data coverage~\cite{shukla1998materialized,
mami2012survey} (e.g., which subset of data this table covers?), data granularity~\cite{rudra2005roles,clark1976effects} (e.g., does each row records one event or an aggregation?), to more domain-specific column understanding~\cite{erickson1992understanding,leech2007array}. 
\end{itemize}

These types of ambiguities present a major challenge in any data task, including Text-to-SQL,
because the LLM needs to contend with both the translation work {\it and} correctly interpreting the query and data semantics. Unfortunately, data ambiguity is relatively unexplored, particularly in the domain of Text-to-SQL.
Prior works study the model sensitivity to query ambiguity  by artificially introducing  ambiguous terms into queries~\cite{wang2022know, zhang2023towards}, and suggest solutions like consulting domain experts via chat interfaces~\cite{yu2019cosql,zhang2020did}. However, these approaches assume clean databases, which is often not true. Additionally,  the user submitting the natural language query is frequently not the data provider and may not understand the subtle assumptions and semantics of the dataset.
In contrast, the data management community has long studied data  quality issues, but mainly in terms of  detecting and fixing data errors~\cite{chu2016data,mahdavi2019raha,rahm2000data} rather  than documenting the dataset in a way that can disambiguate its  application to different data tasks. 
We believe that LLMs, such as GPT-4 with strong general knowledge~\cite{mahowald2023dissociating}, offer an opportunity for data  providers to document their datasets  in natural language. 
While there have been some proposals to better document
datasets~\cite{vardigan2008data,blank2004data,jeng2016toward}, these standards are designed for human understanding and haven't seen wide adoption for general datasets.
Whether LLMs can take advantage of this documentation for data tasks remains an open question.

In this paper, we study how combinations of data and query disambiguation work in isolation and together to improve Text-to-SQL tasks. 
We simulate a scenario where a data  provider documents their data offline, and a user uses natural language to disambiguate their text input online. 
To  delve into ambiguities in real-world datasets, we use KaggleDBQA, a Text-to-SQL benchmark collected from  8 real-world Kaggle databases with 18 tables. This benchmark had annotators draft 272 natural language queries, and SQL experts provided one SQL answer per query. 
KaggleDBQA provides basic data documentation to describe obscure column names. However, data ambiguity in KaggleDBQA goes beyond  obscure column names and encompasses common data ambiguity issues, thus making it an intriguing subject for study. We illustrate these ambiguities with an example.

\vspace{2mm}
\hrule

\begin{example}
\label{exp}
\small
\setstretch{0.25}
Consider the database from KaggleDBQA that records football matches and betting data:
\begin{verbatim}
  betfront: year, datetime, country, competion, match, home opening,..., away closing
  football_data: season, datetime, div, country, league,..., bwd, bwa
\end{verbatim}

Given the natural language query
{\it "Which year has the most matches?"}, there are both query and data ambiguities:
\begin{itemize}[leftmargin=0.5cm]
\item {\bf (Query) Term Ambiguity:~\cite{wang2022know,yu2019cosql,veltri2023data,yu2019cosql,zhang2020did}}
football\_data uses season to represent time and season could span two years. Which year is the query asking for? Is it the start year or the end year?
\item {\bf (Query) Output Schema~\cite{dong2023c3, zhang2023towards}}: What's the expected schema of the output? Is it solely the 'year', or should it also provide the count of matches as the evidence?

\item {\bf (Data) Value Consistency~\cite{madnick2009overview,raman2001potter}}: How are matches formatted? Do they consistently follow the "teamA vs. teamB" format?  This will influence the method of selecting matches.
\item {\bf (Data) Data Converage~\cite{shukla1998materialized,
mami2012survey}}: Do both tables contain every match?  Do they contain mutually exclusive subsets, or do they intersect? This brings up the potential need for union or deduplication operations.
\item {\bf (Data) Data Granularity~\cite{rudra2005roles,clark1976effects}}: Does each row in a table correspond to a unique match, or can there be repeated rows for the same match due to updates in statistics or data as time progresses? This determines if a simple COUNT(*) would suffice or if we need to account for duplications with COUNT(distinct match)
\end{itemize}

Note that the benchmark's provided SQL answer (\texttt{SELECT YEAR FROM betfront GROUP BY YEAR ORDER BY count(*) DESC LIMIT 1}) makes several assumptions to clarify the aforementioned issues: each row in betfront is a unique match, and betfront contains all matches without the need to use football\_data.

\end{example}
\hrule

\section{Disambiguation Methods}
\label{sec:methods}

To control data  and query ambiguity, we introduce methods to disambiguate data and query.

\subsection{Data Disambiguation}

We emulate a situation where data providers offer offline documentation to disambiguate data. We first consider the documentation used by previous works to help LLM understand data:

\begin{itemize}[leftmargin=0.5cm]
\item {\bf Schema}: Previous Text-to-SQL ~\cite{liu2023comprehensive,dong2023c3} and popular open-source projects like Langchain~\cite{Chase_LangChain_2022} and LlamaIndex~\cite{Liu_LlamaIndex_2022} only provide the schema. This can be ineffective for noisy data.
\item {\bf Name Description}: Prior works including KaggleDBQA~\cite{zhang2023data, hegselmann2023tabllm,lee2021kaggledbqa} provide documentation to describe the meanings of the obscure column names.
\item {\bf Sample}: Data samples~\cite{hagleitner2023gpt4sql,moller2023prompt,zhang2021commentary,gao2023text} are provided to help GPT interpret the data. By default, we provide each table's sample of the first 5 rows. 
\end{itemize}

However, none of the above tackle the data ambiguities detailed in \Cref{exp}. 
In response, we provide documentation for three common data ambiguity issues for KaggleDBQA. We disambiguate and  document data by exploring data and interpreting the provided SQL answers:
\begin{itemize}[leftmargin=0.5cm]
\item {\bf Value Consistency~\cite{madnick2009overview,raman2001potter}}: For each column, we document whether the data is represented consistently in some formats. We also specify any outlier formats that exist.
\item {\bf Data Coverage~\cite{shukla1998materialized, mami2012survey}}: We document the coverage of each table, specifying whether it represents the entirety of real-world events or if it has been subsetted in certain ways (e.g., time/location).
\item {\bf Data Granularity~\cite{rudra2005roles,clark1976effects}}: For every table, we document whether the rows represent aggregated data by some group-by keys, or raw data entries for some real-world events.
\end{itemize}

We provided the example documentation for \Cref{exp} in \Cref{doc}.

\noindent{\bf Limitations:}
There are other common types of documentation we've not provided due to challenges in determining the ground truth. For example, half of the columns in "football\_data" have $>30\%$ missing values. It remains ambiguous whether these indicate non-applicability, data collection errors, or censoring~\cite{acock2005working}. KaggleDBQA doesn't address these missing values. In the absence of a reliable ground truth, we abstain from documenting them and leave them for future research.

{\bf Levels of Documentation and Refinement:}
We vary  documentation levels, beginning with the schema and then incrementally incorporating samples, Name descriptions, value consistency, data coverage, and data granularity. However, we observed that certain documentations are repetitive. Naively adding more  adversely impacts GPT-4, because lengthy and irrelevant prompts hinder the LLM from focusing on the useful information~\cite{liu2023lost,daniluk2017frustratingly}. 
Therefore, we refine documentation in two ways:
(1) Documentation describing each column (name description and value consistency) tends to be lengthy. We employ an agent approach~\cite{Liu_LlamaIndex_2022, Chase_LangChain_2022} first to let GPT-4 select up to 5 columns. Then, documentation is provided only for these.
(2) Name description, sample, and value consistency have many overlaps, as they similarly help GPT understand the columns. We therefore only provide one.

\begin{table}
  \caption{Types of documentations for Data Disambiguation, and Query Disambiguation Methods}
  \label{doc}
  \centering
  \begin{tabular}{p{3cm}p{10cm}}
    \toprule
    {\bf Type }    & {\it Example}    \\
    \midrule
    {\bf Name Description} & {\it "bwd means Bet\&Win draw odds."}      \\
    % \midrule
    {\bf Value Consistency}     & {\it "Matches are consistently denoted in the format of 'home team - away team', for example, 'Malta - Albania'. There are no outliers."}    \\
    {\bf Data Converage}     & {\it 'football\_data' covers all the matches only from 2009-2013."}   \\
    {\bf Data Granularity}  & {\it "Each row in 'betfront' reports for each unique match in each competition, the detailed time, location and betting records. It is not aggregated."}    \\

\midrule
    {\bf Term Ambiguity}     & {\it "In which year did the most matches take place?"}    \\
    {\bf Output Schema}     & {\it "The output must only contain the year."}    \\
    \bottomrule
  \end{tabular}
  \vspace{-5mm}
\end{table}

\subsection{Query Disambiguation}

We disambiguate queries in two ways: (1) {\bf Output Schema:}
Almost all queries within KaggleDBQA have underspecified output schemas~\cite{dong2023c3, zhang2023towards}. We explicitly specify the output schema for all queries. (2) {\bf Term Ambiguity:} Some queries also have ambiguous terms~\cite{wang2022know,yu2019cosql,veltri2023data,yu2019cosql,zhang2020did}. For instance, some queries ask about the "most dangerous places" without explaining what "dangerous" means. Other queries ask for crimes in  "Manchester", which could refer to Greater Manchester or the city of Manchester. We carefully  review each query and refine these terms based on the provided answers.

\section{Experiments}

{\bf Data and Model:} 
Due to the manual nature of disambiguation, we evaluate 2 (out of 8) KaggleDBQA databases (Soccer and Crime) with 45 queries. 
We use GPT-4 model with 8K context size. We employ the standard chain-of-thought 
to enhance GPT's performance and interpretability~\cite{wei2022chain}.

{\bf Evaluation Setting:} 
In line with previous studies~\cite{lee2021kaggledbqa,yu2018spider}, we assess SQL queries using exact match accuracy. 
It's possible for GPT-4 to produce semantically the same  but syntactically different queries; we manually evaluate the queries to ensure that this doesn't occur.
Despite our efforts to resolve ambiguities using the provided SQL answers (\Cref{sec:methods}), we observed that 22.2\% of them have errors:
(1) {\it 13.3\% have inconsistencies}: We note variations in the interpretation of  terms. E.g., for queries asking "area", some answers use the "lsoa" column (Lower Layer Super Output Areas), while others opt for the "location" (street-level). For consistency, we fix SQL answers to be consistent with the "lsoa" interpretation.
(2) {\it  8.9\% have syntax errors}: We detect errors of missing 'Distinct' in count and improper null checks (= "" instead of "is Null"). We  fix these syntax errors in SQL answers.

{\bf Error Analyses:} 
We highlight two common mistakes GPT-4 made: (1) Adding extra columns to the output(\texttt{Output})~\cite{dong2023c3}. (2) Using exact string matches instead of fuzzy ones for selection (\texttt{Fuzzy}). 
If the only mistake GPT-4 makes falls into these two categories, we'll highlight them. 
For any other errors or if there are additional mistakes, we label them as \texttt{Other}.

\begin{figure}[t]
  \centering
  \begin{subfigure}[b]{\linewidth} % Define the width of the subfigure as a fraction of the line width
    \centering
    \includegraphics[width=1\linewidth]{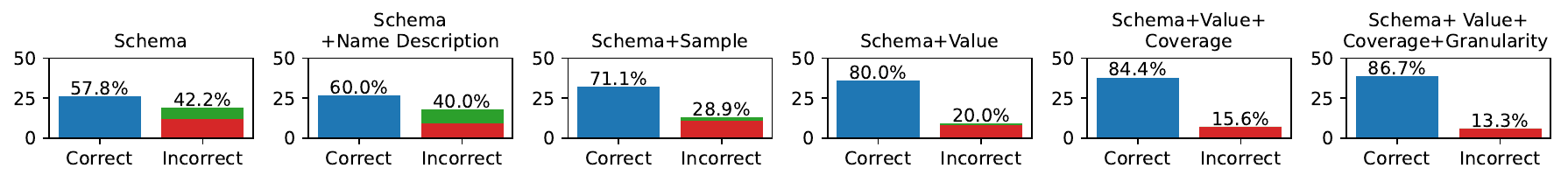}
    \vspace{-5mm}
    \caption{Accuracy when the queries have been disambiguated, but the levels of documentation vary.}
    \label{fig:output_plot1}
  \end{subfigure}
  \hfill % Fill the horizontal space
  \begin{subfigure}[b]{\linewidth}
    \centering
    \includegraphics[width=1\linewidth]{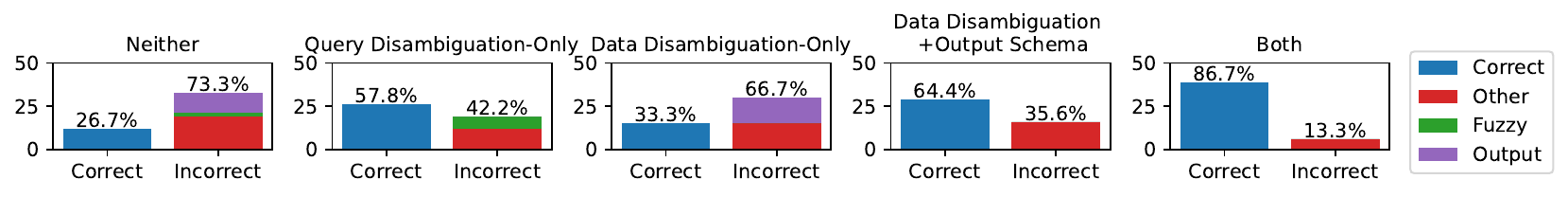}
     \vspace{-5mm}
    \caption{Accuracy when the queris and data are disambiguated in isolation or together.}
    \label{fig:output_plot2}
  \end{subfigure}
  \caption{GPT-4 Error analysis. Blue bars are for correct, while others are for distinct types of errors.}
  \label{fig:output_plots}
  \vspace{-5mm}
\end{figure}

\subsection{Levels of Documentation for Data Disambiguation}

We first disambiguate queries, and study how varying levels of documentation help Text-to-SQL.

{\bf Results:} \Cref{fig:output_plot1} shows the results. (1) We find that GPT-4  achieves a high accuracy of  $57.8\%$ with only schema. Most errors arise due to the preference for exact string matching over fuzzy matching.   (2) In contrast, only a $2.2\%$  improvement is observed when \texttt{Name Description} is provided. We find that GPT-4 has the capability to infer full names from vague column names. KaggleDBQA documentation doesn't significantly aid GPT-4.  (3) Giving samples helps GPT-4 avoid most \texttt{Fuzzy} errors. (4) By providing documentation on Value Consistency, GPT-4 can better avoid \texttt{Fuzzy} and apply correct predicates. E.g., with only samples, GPT-4 misinterprets "season" in $football\_data$ as only in the format of "Year1/Year2". By providing documentation specifying that
"season" also contains a single year, GPT-4 fixes selection errors.   (5)  Data Coverage helps GPT-4 avoid mistakes in unioning and joining the "betfront" and "football\_data" tables,  as it understands a single table is sufficient for the query. (6) Data Granularity assists GPT-4 in applying predicates. E.g., one query asks for "street" crimes. GPT-4 previously misunderstood one row as one street crime. By specifying the row granularity as a crime on streets, roads or avenues, GPT-4 correctly refines the selection.

\subsection{Compare Query vs Data Disambiguation}

We assess the effects of query disambiguation (original, or with term and output schema disambiguated) versus data disambiguation (schema only, or schema with value, coverage, and granularity).

{\bf Results:} \Cref{fig:output_plot2} shows the results. (1) When only the schema and the original query are provided, the accuracy is $26.7\%$ and matches the KaggleDBQA results. (2) Query Disambiguation is pivotal: Replacing the original queries with disambiguated ones elevates the accuracy to $57.8\%$. (3) If we only disambiguate data but not query, accuracy is only $33.3\%$.While this might suggest that data disambiguation by itself isn't as effective, we discover that $33.3\%$ of the errors come from the output schema, which is easy to fix. (4) To verify this, we specify output schema (not term disambiguation) for queries, and the accuracy surges to $64.6\%$. This underlines the importance of data disambiguation. (5) Finally, disambiguating both yields the highest accuracy at $86.7\%$.  $13.3\%$ errors remain even with both documentation and query disambiguation. We investigate these and find that they are from domain-specific nuances. E.g., "home losing odds" corresponds to "away winning odds", but GPT-4 chooses "home winning odds". Addressing these needs domain-specific documentation~\cite{zhang2023sciencebenchmark}.

\section{Conclusion}

This work studies ambiguities in real-world datasets and assesses how documentation aids GPT-4 in enhancing Text-to-SQL.
Our findings reveal that data ambiguities are prevalent, and extend beyond obscure column names to issues like  value consistency, data coverage, and granularity. 
By providing documentation on these issues, GPT-4’s accuracy is improved by $28.9\%$.
Looking forward, we intend to (1) investigate other data ambiguity issues, such as missing values, and (2) explore semi-automating the documentation process  by leveraging GPT-4 to assist data providers.

\section*{Acknowledgements}

This work was funded by the NSF under Grant Numbers 1845638, 2008295, 2106197, 2103794, 2312991, and was further supported by the Google PhD Fellowship, along with contributions from Amazon and Adobe.
\bibliographystyle{plainnat}
\bibliography{references}

\end{document}